# Measuring the string susceptibility in 2D simplicial quantum gravity using the Regge approach

*Christian Holm*[1] and *Wolfhard Janke*[1,2]

[1] Institut für Theoretische Physik, Freie Universität Berlin
Arnimallee 14, 14195 Berlin, Germany

[2] Institut für Physik, Johannes Gutenberg-Universität Mainz
Staudinger Weg 7, 55099 Mainz, Germany

**Abstract**

We use Monte Carlo simulations to study pure 2D Euclidean quantum gravity with $R^2$-interaction on spherical topologies, employing Regge's formulation. We attempt to measure the string susceptibility exponent $\gamma_{\rm str}$ by using a finite-size scaling Ansatz in the expectation value of $R^2$, as has been done in a previous study by Bock and Vink (Nucl. Phys. B438 (1995) 320). By considerably extending the range and statistics of their study we find that this Ansatz is plagued by large systematic errors. The $R^2$ specific string susceptibility exponent $\gamma'_{\rm str}$ is found to agree with theoretical predictions, but its determination also is subject to large systematic errors and the presence of finite-size scaling corrections. To circumvent this obstacle we suggest a new scaling Ansatz which in principle should be able to predict both, $\gamma_{\rm str}$ and $\gamma'_{\rm str}$. First results indicate that this requires large system sizes to reduce the uncertainties in the finite-size scaling Ansätze. Nevertheless, our investigation shows that within the achievable accuracy the numerical estimates are still compatible with analytic predictions, contrary to the recent claim by Bock and Vink.

# 1 Introduction

Two-dimensional (2D) Euclidean quantum gravity is believed to be an important toy model on our way to a realistic quantum theory of gravity. Along with the theoretical achievements of string theories 2D Euclidean quantum gravity has become a rather well understood subject. In contrast to the classical theory, which is dynamically trivial, the quantum theory possesses a rather rich structure due to the conformal anomaly. One of the interesting aspects is the intrinsic fractal structure of space time which shows up in the divergence of the partition function $Z(A)$ with increasing area $A$, governed by the string susceptibility exponent $\gamma_{\rm str}$ [1]:

$$Z(A) \propto A^{\gamma_{\rm str}-3} e^{-\lambda_R A}. \qquad (1)$$

Here $\lambda_R$ denotes the renormalized cosmological constant. The exponent $\gamma_{\rm str}$, which depends on the genus $g$ of the surface, has been calculated first by Knizhnik, Polyakov, and Zamalodchikov (KPZ), using conformal field theory methods [2] to be

$$\gamma_{\rm str} = 2 - \frac{5}{2}(1-g). \qquad (2)$$

Because $\gamma_{\rm str} = 2$ is the classically expected result, quantum effects can only be seen for non-toroidal topologies ($g \neq 1$). For spherical topologies ($g = 0$) the prediction $\gamma_{\rm str} = -1/2$ agrees with matrix model methods [3]. Moreover, if one couples spin matter to gravity, conformal field theory and matrix models both predict a significant change in the critical exponents of the associated spin phase transition when compared to flat space exponents [2, 4]. These remarkable agreements strengthened the belief in the existence of universal features of 2D quantum gravity.

The matrix model approach gave rise to a novel numerical scheme, called the dynamically triangulated random surface (DTRS) approach to quantum gravity [5]. The DTRS method could reproduce the prediction for $\gamma_{\rm str}$ [1] as well as the critical spin exponents [6].

For the alternative numerical approach using Regge calculus [7, 8] the situation is far away from being clear. One major conceptual problem is associated with the path integral measure of quantum gravity (for a good review see Ref. [9]). There have been almost as many different proposals for the integration measure as there are different approaches to quantum gravity. Within the gauge theory approach, depending on one's own preference in the correct gauge group of gravity, such as diffeomorphism, local Lorentz, Poincaré or conformal group, one obtains different results for the integration measure. The same situation holds true if one uses the Hamiltonian formalism or tries to approach the problem by measure theory [9]. Most measures differ by the power of the determinant of the metric $g$, which stands in front of the integration measure, but also non-local objects of the Faddeev-Popov type can appear.

We would like to stress that this ambiguity resides in the continuum approach to quantum gravity, and hence should haunt any quantum theory of gravity; it is not a problem confined to Regge calculus. In light of the agreement of conformal field theory and matrix models, however, one is inclined to think that at least in two dimensions there might exist something like a unique measure and that both, the KPZ and the matrix model approach, capture all the essential features of this measure to fall into the same universality class. To support this argument one would like to see Regge calculus as a third unrelated candidate for a quantum theory of gravity to reproduce all known results (for a different opinion, however, see Refs. [10, 11]). This would enforce the hope that also in higher dimensions the details of the numerical approach are not overly important.

Unfortunately the results of numerical investigations using Regge calculus have been disappointing so far. Using the commonly employed $dl/l$ measure on a Regge lattice, no change in the phase transition of an Ising model coupled to gravity was observed [12, 13], the critical exponents remained in the flat space Onsager universality class. Still there is the hope that with a different measure or a different spin coupling to gravity one can reproduce the KPZ critical exponents [14, 15]. For pure gravity and toroidal topology agreement with $\gamma_{\text{str}} = 2$ was reported [12, 16], but, as noted before, the torus is not a good testing ground for (2). For the sphere and the $dl/l$ measure, Gross and Hamber [12] found weak numerical evidence for $\gamma_{\text{str}} = -1/2$, while Bock and Vink in Ref. [16] clearly did not see the dependence of (2). Soon afterwards, however, it was claimed in Ref. [17] that the string susceptibility exponent $\gamma'_{\text{str}}$ for strong $R^2$-gravity was consistent with the theoretical predictions [18] for the sphere ($g = 0$), but not for the bi-torus ($g = 2$). In light of these contradicting results, we found it necessary to reinvestigate this problem again.

In the Regge approach part of the difficulty to estimate $\gamma_{\text{str}}$ is the lack of methods to measure $\gamma_{\text{str}}$ directly. All approaches used so far introduce a curvature square term $R^2$ and deduce from its expectation value an estimate on $\gamma_{\text{str}}$ through a finite-size scaling analysis. The problem with $R^2$-gravity is that there are actually two distinct scaling regimes, each of which defines its own string susceptibility. For the model defined by

$$Z(A) = \int \mathcal{D}\mu(g) e^{-S_G} \delta(\int d^2x \sqrt{g} - A), \qquad (3)$$

with the gravitational action taken as

$$S_G = \int d^2x \sqrt{g}(\lambda + \frac{a}{4}R^2), \qquad (4)$$

and $\mathcal{D}\mu(g)$ being the DeWitt measure, the scaling behavior of $Z(A)$ was investigated by Kawai and Nakayama [18]. The coupling constant $a$ has the dimensions of an area, and therefore sets a length scale of $\sqrt{a}$. The dimensionless quantity $\hat{A} := A/a$ can then be used to distinguish between the cases of weak $R^2$-gravity ($\hat{A} \gg 1$) and strong $R^2$-gravity ($\hat{A} \ll 1$). For the case $\hat{A} \gg 1$ the scaling relations (1) and (2) are recovered, whereas for $\hat{A} \ll 1$ it was found, that

$$Z(A) \propto A^{\gamma'_{\text{str}} - 3} e^{-S_c/\hat{A}} e^{-\lambda_R A - \xi \hat{A}}, \qquad (5)$$

where
$$\gamma'_{\text{str}} = 2 - 2(1-g), \qquad (6)$$

$S_c = 16\pi^2(1-g)^2$ is the classical action, and $\xi$ is some constant. Only for the torus ($g = 1$) the two scaling behaviors and string susceptibilities are the same. Note that the dependence of $\gamma'_{\text{str}}$ on the gender of the surface is, loosely speaking, weaker than that of $\gamma_{\text{str}}$. For later use we also write down the derivative of $Z$ with respect to $\hat{A}$:

$$\frac{\partial \ln Z}{\partial \hat{A}} = \frac{\gamma_{\text{str}} - 3}{\hat{A}} - a\lambda_R \qquad (\hat{A} \gg 1), \qquad (7)$$

$$\frac{\partial \ln Z}{\partial \hat{A}} = \frac{\gamma'_{\text{str}} - 3}{\hat{A}} - a\lambda_R + S_c/\hat{A}^2 - \xi \qquad (\hat{A} \ll 1). \qquad (8)$$

While the partition function $Z$ is not directly accessible in Monte Carlo simulations, logarithmic derivatives like (7) and (8) can be estimated by measuring appropriate expectation values.

The remainder of the paper is organized as follows. In Sec. 2 we first review the Regge discretization method. Then we discuss the finite-size scaling methods used in Ref. [16, 17] to determine $\gamma_{\rm str}$ and $\gamma'_{\rm str}$, show its shortcomings, and suggest a new method to measure $\gamma_{\rm str}$ and $\gamma'_{\rm str}$. Section 3 deals with some details of our Monte Carlo simulations. In Sec. 4 we present our numerical results and discuss their interpretation, and in Sec. 5 we conclude with a brief summary of the main results and some final remarks.

## 2 Discretization Method and Finite-Size Scaling Ansätze

### 2.1 Regge discretization

Regge's discretization program [7] consists of replacing a given continuum manifold by piecewise linear manifolds, whose internal geometry is flat. This procedure works for any space-time dimension and for metrics of arbitrary signature. Here we restrict ourselves to the simplest case of two dimensions and Euclidean signature.

In two dimensions this procedure is most easily visualized by choosing a triangulation of the surface under consideration, where each triangle then represents a piecewise linear manifold. The net of triangles is itself a two geometry, with singular (non-differentiable) points located at the vertices of the net, where several triangles meet. A vector that is linearly transported around these vertices experiences in the presence of curvature a rotation by the deficit angle $\delta_i = 2\pi - \sum_{t \supset i} \theta_i(t)$, where $\theta_i(t)$ is the dihedral angle at vertex $i$. For the area assignment we used the barycentric decomposition, where $A_i = \sum_{t \supset i} A_t/3$ denotes the barycentric area with $A_t$ being the area of a triangle $t$. We can then identify the following continuum quantities with their discrete counterparts, namely

$$\int d^2x \sqrt{g(x)} \quad \longrightarrow \quad \sum_i A_i, \qquad (9)$$

$$\int d^2x \sqrt{g(x)} R(x) \quad \longrightarrow \quad 2\sum_i \delta_i, \qquad (10)$$

$$\int d^2x \sqrt{g(x)} R^2(x) \quad \longrightarrow \quad 4\sum_i \frac{\delta_i^2}{A_i}. \qquad (11)$$

Our notation is identical to that used in Ref. [13]. In two dimensions the Einstein-Hilbert action $\int d^2x\sqrt{g(x)}R(x)$ is by the Gauss-Bonnet theorem a topological invariant, which makes such a theory classically trivial, there are no equations of motion. Regge [7] gave a beautiful proof of this theorem in terms of the deficit angle. The sum over the deficit angles in two dimensions is proportional to the Euler characteristic, namely $\sum_i \delta_i = 4\pi(1-g)$. We therefore will not consider the corresponding term in the gravitational action. If we keep the area $A$ fixed to its initial value, then, classically, dynamics can only arise from the $R^2$-interaction term. Such a term was used in two and higher dimensional studies to cure the unboundedness problem of the gravitational action [19].

For triangulated surfaces the Euler relation reads as

$$N_0 - N_1 + N_2 = 2(1-g), \qquad (12)$$

where $N_0, N_1$, and $N_2$ denote the number of sites, links and triangles, respectively. For triangulations without boundary we also know that a link is shared by two triangles, resulting in the relation $N_1/3 = N_2/2$. From these two relations one can derive two more, namely $N_0 - 2(1-g) = N_2/2$ and $N_0 - 2(1-g) = N_1/3$, which will become useful later.

For each triangle there is a one-to-one correspondence between the square of the link lengths and the components of the metric. Denoting by $g_{\mu\nu}(i)$ the components of the metric tensor for the $i^{th}$ triangle, and by $q_{i+\mu,i+\nu}, q_{i,i+\mu}$, and $q_{i,i+\nu}$ the square of its three edge lengths, one can derive the following relation $g_{\mu\nu}(i) = \frac{1}{2}[q_{i,i+\mu} + q_{i,i+\nu} - q_{i+\mu,i+\nu}]$. In classical Regge calculus one starts with the action principle and derives the equations of motion, one for each link. The link lengths have to be adjusted to satisfy those equations in order to be a classical solution. The connectivity of the edges, in simplicial topology called the incidence matrix, is fixed from the beginning through the simplicial decomposition of the manifold under consideration.

In quantum Regge calculus the technical aspects are similar, although the philosophy is quite different. Here we want to evaluate the functional integral in Eq. (3) by Monte Carlo (MC) methods. In principle, the integral has to be extended over all metrics on all possible topologies, but, as usual, we restrict ourselves to a specific topology, here the sphere. The integral over the metric is replaced by an integral over the square of the link lengths. An important ingredient in the functional-integral method is the appropriate measure, which, as already explained in the Introduction, is not even known in the continuum. The most popular measure is DeWitt's supermetric [20], a distance functional on the space of metrics. It was used by Polyakov in his famous string solution [21]. Because in 2D the measure is the primary source of the non-trivial dynamical content of the theory, its correct transcription might be the key point in reproducing the KPZ results. Nevertheless if the discretized DeWitt measure is still a local one, then one might argue on the basis of universality that any other local measure will do as well. In the present study we only report simulations with the most commonly used "computer" measure $\mathcal{D}\mu(l) = \prod_{\langle ij \rangle} \frac{dl_{ij}}{l_{ij}} F(l_{ij})$, in order to facilitate a comparison of our data with the results of Ref. [16]. Here $F(l)$ is a function which takes the value one, if the links $l_{ij}$, forming one triangle, fulfill the triangle inequalities, and zero otherwise. For small link lengths we introduced a scale invariant cut-off to avoid round-off problems. We will take up the question of the measure dependence of $\gamma_{\text{str}}$ ($\gamma'_{\text{str}}$) in a forthcoming publication [22].

Collecting the transcriptions from the continuum to the simplicial Regge approach, our lattice analogue to Eq. (3) is therefore given by

$$Z(A, N_1) = \int \prod_{\langle ij \rangle} \frac{dl_{ij}}{l_{ij}} F(l_{ij}) e^{-\sum_i (\lambda A_i + a R_i^2)} \delta(\sum_i A_i - A), \qquad (13)$$

where we abbreviated $R_i^2 := \delta_i^2 / A_i$.

## 2.2 Finite-size scaling

The method to extract $\gamma_{\text{str}}$ through the finite-size scaling (FSS) properties of the expectation value of an added $R^2$-interaction term was first discussed by Gross and Hamber [12], and later improved in Ref. [16]. Here we would like to point out that there is still room for ambiguities in the FSS assumptions. A very simple derivation of the scaling behavior comes from a rescaling argument. We will not discuss in this paper any measure ambiguities, and restrict ourselves to the scale invariant computer measure $dl/l$, as used in our (and most previous) simulations. Other measures could be treated without any problem. The simple derivation presented below has the additional advantage over the one given in Ref. [16], that it neither invokes Ward identities nor any artificial weight functions.

Consider the partition function (13). A rescaling of the link lengths $l$ to the dimensionless link lengths $l'$ of the form $l \longrightarrow l'\sqrt{A}$ yields $\sum_i A_i \longrightarrow A \sum_i A'_i$, and therefore

$$Z(A, N_1) = \int \prod_{\langle ij \rangle} \frac{dl'_{ij}}{l'_{ij}} F(l'_{ij}) e^{-A \sum_i (\lambda A'_i + \frac{a}{A^2} R'^2_i)} \frac{1}{A} \delta(\sum_i A'_i - 1). \qquad (14)$$

From this expression one can easily compute the derivative

$$\frac{\partial \ln Z}{\partial \hat{A}} = -a\lambda + \frac{1}{\hat{A}}(\hat{R}^2 - 1), \qquad (15)$$

where

$$\hat{R}^2 := a \langle \sum_i R_i^2 \rangle = a \langle \sum_i \delta_i^2 / A_i \rangle, \qquad (16)$$

and by inspecting (14) it is also easy to see, that $\hat{R}^2 = \hat{R}^2(\hat{A}, N_1)$, meaning that the expectation value depends only on $N_1$ and the dimensionless parameter $\hat{A}$. For large $N_1$ one expands the finite part of $\hat{R}^2$ into a power series, whose first three terms read as

$$\hat{R}^2 = \ldots + b_0 \hat{A} + b_1 + b_2/\hat{A} + \ldots . \qquad (17)$$

If this is inserted in (15), then a comparison with (7), (8) gives $b_0 = -(\lambda_R - \lambda)a, b_1 = \gamma_{\text{str}} - 2$ for $\hat{A} \gg 1$, and $b_0 = -\xi - (\lambda_R - \lambda)a, b_1 = \gamma'_{\text{str}} - 2, b_2 = S_c$ for $\hat{A} \ll 1$.

The scaling Ansatz of Refs. [16, 17] is to consider first a power-series expansion of $\hat{R}^2(\hat{A}, N_1)$ in $N_1$. They actually use $N_2$ instead of $N_1$, but this is trivial because for any compact triangulation we have the fundamental relation $3N_2 = 2N_1$. Furthermore, their expansion of $\hat{R}^2$ is not done at fixed $\hat{A}$, but at a fixed discretization scale set by the average triangle area $a_0 := A/N_2$ through the dimensionless parameter $\hat{a}_0 := a_0/a$:

$$\hat{R}^2(\hat{a}_0, N_2) = N_2 c_0(\hat{a}_0) + c_1(\hat{a}_0) + c_2(\hat{a}_0)/N_2 + \ldots . \qquad (18)$$

The coefficients $c_i(\hat{a}_0)$ are thus defined in the thermodynamic (infinite area) limit. This expansion has to be justified by the simulation results, because it is not based on any *ab initio* calculations. A similar good educated guess would have been to to consider for instance an expansion in a linear length parameter $\sqrt{N_2}$. In a second step the coefficients $c_i$ are expanded in Refs. [16, 17] into a power series in $\hat{a}_0$ as

$$c_0 = c_0^{(0)} + \hat{a}_0 c_0^{(1)} + \ldots, \tag{19}$$

$$c_1 = c_1^{(0)} + \hat{a}_0 c_1^{(1)} + \ldots, \tag{20}$$

$$c_2 = (c_2^{(0)} + \hat{a}_0 c_2^{(1)} + \ldots)/\hat{a}_0, \tag{21}$$

and the continuum limit is taken by sending the discretization scale to zero, $\hat{a}_0 \to 0$. A comparison with (17) then yields $b_0 = c_0^{(1)}$, $b_1 = c_1^{(0)}$, and $b_2 = c_2^{(0)}$. Note, that in order to make contact with the continuum result of Eq. (8), $c_2$ needs to start with a divergent term $1/\hat{a}_0$. Only in the combined limit $N_2 \to \infty, \hat{a}_0 \to 0$, this makes sense. But because $\hat{a}_0$ is fixed, and not $\hat{A}$, effectively there is no control of the crossover from strong $R^2$-gravity scaling behavior ($\hat{A} \ll 1$) to the weak $R^2$-gravity scaling behavior ($\hat{A} \gg 1$). If one takes first the thermodynamic limit in (18) then one always obtains the values of the coefficients $c_i$ in the limit $\hat{A} = N_2 \hat{a}_0 \gg 1$, hence for weak $R^2$ gravity. This means, $c_1^{(0)} = \gamma_{\text{str}} - 2$, and $c_2 \to 0$, but one has to be careful to make the system size always large enough to reach this limit. If one truncates the fit at some suitable value of $N_2$ to explore the region where $\hat{A} \ll 1$, then the continuum limit can only be taken at finite $N_2$. Because the results of Ref.[17] were obtained in for very small $N_2$ finite-size effects can become important. Even worse, because the coefficients $c_i^{(j)}$ in the expansion (20) are constants, how can $c_1^{(0)} + 2$ change from $\gamma_{\text{str}}$ to $\gamma_{\text{str}}'$, as is claimed in Ref. [17]? These subtleties, which have not been previously addressed, make it appear very unlikely that one can unambiguously extract $\gamma_{\text{str}}'$ in this way.

We therefore suggest an alternative approach, where we look at the FSS behavior at a constant value of $\hat{A}$. This is very much in the spirit of the investigations of Ref. [18]. Expanding $\hat{R}^2(\hat{A}, N_2)$ at constant $\hat{A}$ we obtain

$$\hat{R}^2(\hat{A}, N_2) = N_2 d_0(\hat{A}) + d_1(\hat{A}) + d_2(\hat{A})/N_2 + \ldots. \tag{22}$$

For large $\hat{A}$ we have to increase $N_2$ to insure that the following terms remain small. Also here other FSS Ansätze are possible. The next step is to expand the coefficients $d_i$ as a power series in $\hat{A}$. This time the coefficient $d_1$ carries all the necessary information to extract the string susceptibilities. A comparison with (17) yields

$$d_1(\hat{A}) = b_0 \hat{A} + (\gamma_{\text{str}} - 2) + \mathcal{O}(1/\hat{A}) \quad \text{for } \hat{A} \gg 1, \tag{23}$$

$$d_1(\hat{A}) = S_c/\hat{A} + (\gamma_{\text{str}}' - 2) + b_0 \hat{A} + \mathcal{O}(\hat{A}^2) \quad \text{for } \hat{A} \ll 1. \tag{24}$$

If we plot $d_1$ versus $\hat{A}$ we thus expect to see a linear behavior for very large $\hat{A}$ and a divergent behavior for small $\hat{A}$, governed by the classical action $S_c = 16\pi^2(1-g)^2$. The difference between our method and that of Ref. [16] appears as a subtle interchange of thermodynamic and continuum limit. We first take the continuum limit ($N_2 \longrightarrow \infty$) for fixed $\hat{A}$, and then the thermodynamic limit, whereas in (18) first the thermodynamic limit is taken for fixed $\hat{a}_0 = \hat{A}/N_2$, and then the continuum limit is performed.

The appearance of the classical action is not hard to understand. Actually, for any regular triangulation with coordination number $q$ of arbitrary topology we find from the Euler relation $\delta_i = 4\pi(1-g)/q$, and $A_i = A/q$, therefore $aR^2 = a\sum \delta_i^2/A_i = 16\pi^2(1-g)^2/\hat{A}$. For small $\hat{A}$ this will be the dominant term.

Let us assume that we have a distribution $q_{ij}^{(0)}$ of the square of the link lengths that give the classical action $S_c = S(q_{ij}^{(0)})$. For small $\hat{A}$ one should be able to make a stationary (semi-classical) approximation of the action $S(q)$ around $S(q^{(0)})$ in the variables $q$, which should become small in the continuum limit due to the constancy of $A$. Our results suggest that it looks like

$$S(q) \approx S_c + N_0 F(\hat{A}) + \sum_{<ij><kl>} S''(q^{(0)})(q_{ij} - q_{ij}^{(0)})(q_{kl} - q_{kl}^{(0)}) + \ldots, \quad (25)$$

where the primes denote differentiation with respect to the $q$, and $S'(q^{(0)}) = 0$. $F$ denotes the first correction to $S_c$ independent of $q$, which is responsible for the divergent behavior of $\hat{R}^2$ proportional to $N_0$. This form of (25) would heuristically explain how $d_0$ and the classical action part in $d_1$ arises. To zeroth order always the classical part dominates, and it is only in the limit of large $\hat{A}$ that quantum fluctuations become more important. If one could compute the second derivative part of $S(q)$, then one could probably also calculate $\gamma'_{\text{str}}$.

Notice that if one uses $N_0$ instead of $N_2$ in Eqs. (18) and (22), one will get in general different coefficients $c_i$ and $d_i$, respectively. An expansion of

$$\hat{R}^2(\hat{a}_0, N_0) = N_0 c_0^* + c_1^* + c_2^*/N_0 + \ldots \quad (26)$$

instead of (18) would give for spherical topologies $c_0^* = 2c_0$, $c_1^* = c_1 - 4c_0$, and $c_2^* = c_2/2$. A further extrapolation of the coefficients $c_i$ to $\hat{a}_0 \to 0$ would therefore yield the same string susceptibility only if $c_0$ vanishes in this limit.

This is not supported by our data, where we see that $c_0$ actually increases as $\hat{a}_0$ decreases. For the case of Eq. (22) the situation is similar. The use of $N_0$ instead of $N_2$ results in $d_0^* = 2d_0$, $d_1^* = d_1 - 4d_0$, and $d_2^* = d_2/2$. This results in a decrease of the string susceptibilities by $4d_0$. We stress that we believe that, a priori, $\hat{R}^2$ should be considered as a function of $N_1$, and that it should therefore be expanded in $N_1$. Because the use of $N_2$ is just a simple scale change, this would be equally valid. Still it is troublesome that even in the continuum limit, the difference between the expansions using $N_2$ and $N_0$ remains. From the Euler formula one can see how the difference depends for general topologies on the gender of the surface. The only changes are $c_1^* = c_1 - 4(1-g)c_0$ and $d_1^* = d_1 - 4(1-g)d_0$. Again only for the torus ($g = 1$) this ambiguity is absent.

We attribute above ambiguity in the determination of $\gamma'_{\rm str}$ ($\gamma_{\rm str}$) to the fact that $\gamma'_{\rm str}$ ($\gamma_{\rm str}$) comes from a sub-leading effect, whereas the leading effect $Z \propto e^{-S_c/\hat{A}}$ can be determined with both, $N_0$ and $N_2$. For Eq. (18) this can be seen by simply inserting the definitions of the coefficients. One then finds $c_2^{(0)*} = \frac{1}{2} c_2^{(0)}$. For Eq. (22) one needs to assume that $d_0(\hat{A})$ has no $1/\hat{A}$ dependence for $\hat{A} \ll 1$.

## 3 The Simulation

All MC simulations were performed on spherical topologies, which were realized as triangulated surfaces of a three-dimensional cube, as discussed in Ref. [16]. For this choice of discretization six vertices have coordination number four, whereas the rest has coordination number six, see Fig. 1. The number of vertices $N_0$ is related to the edge length $L$ of the cube by $N_0 = 6(L-1)^2 + 2$. Usually, the size of the lattices varies from $L = 7$ up to 55, corresponding to 218 up to 17498 lattice sites, or 648 up to 52488 link degrees of freedom. To update the links we used a standard multi-hit Metropolis update with a hit rate ranging from $1 \ldots 3$. In addition to the usual Metropolis procedure a change in link length is only accepted, if the links of a triangle fulfill the triangle inequalities.

The area $A$ was kept fixed at its initial value $A = \sum_i A_i = N_2/2$ during the update to simulate the delta function in Eq. (13). In order to achieve this, we would need in principle to rescale all links during each link update, amounting in a non-local procedure. However, due to the scaling properties of the partition function, this can be absorbed in a simple scale factor in front of the $R^2$ term. To avoid round-off errors we explicitly performed a rescaling after every full lattice sweep. Notice, that technically our simulation procedure is different from the methods employed in Refs. [16, 17].

The first set of simulations was designed to test the extrapolations to the continuum limit which in Ref. [16] were based on very small values of $1/\hat{a}_0$ in the range of $0.5 - 5$. Already exploratory runs [23] indicated that this range is not appropriate to see the true asymptotic behavior. In the present study we therefore used much larger values of $1/\hat{a}_0 = 0.2, 2.5, 10, 20, 40, 80, 160, 320, 640,$ and $1280$.

The second set of simulations consisted of runs at constant $\hat{A}$ at the values of $\hat{A} = 9126/a$ with $a = 1.25, 2.5, 5, 10, 20, 40, 80, 160, 320,$ and $640$, which covers roughly the range of $\hat{A} = 14 - 7300$.

For each run we recorded about 10000 measurements of the curvature square $R^2 = \sum_i \delta_i^2 / A_i$. To reduce temporal correlations, we took measurements every second MC sweep through the entire lattice. The statistical errors were computed using standard jack-knife errors on the basis of 20 blocks. The integrated autocorrelation time $\tau_{R^2}$ of $R^2$ was usually in the range of 5 - 10.

## 4 Results

### 4.1 Scaling at fixed $\hat{a}_0$

We first take a look on the raw simulation data, and plot $\hat{R}^2/N_2$ versus $1/N_2$. The set of data points for the four largest values of $\hat{a}_0$ can be inspected in Fig. 2. Because we chose $\hat{a}_0 = 1/2a$ this is equivalent to the simulations with the four smallest values of $a$. The curves look straight to the eye, and the scaling Ansatz (18) seems to works well even without the $c_2$ coefficient, indicating that $N_2$ is large enough so that we are in the weak $R^2$-gravity regime.

A closer look on the curves for $1/\hat{a}_0 = 10$ ($1/\hat{a}_0 = 20$), see Fig. 3, shows apparently two scaling regions, divided approximately by a line through $1/N_2 \approx 0.005(0.003)$. We interpret this region as the crossover region from $\hat{A} \gg 1$ to $\hat{A} \ll 1$. Because $\hat{A}$ was chosen to be $N_2/2a$, we decrease $\hat{A}$ either by decreasing $N_2$ or by increasing $a$. This means we always start out on small lattices in the strong $R^2$-gravity regime, and end up on sufficiently large lattices always in the weak $R^2$-gravity regime. It is therefore hard to imagine, that one can fit the whole range of data points with the same truncated Ansatz (18) without taking into account contributions from other coefficients which arise from interchanging the $\hat{A}$ limits. We know the scaling behavior only for the two limiting cases of $\hat{A}$, but nothing in between these two limits. By increasing $N_2$ we would again be able to extract $c_1$ with a linear Ansatz, yielding $\gamma_{\text{str}}$, but already at the rather coarse discretization scale of $\hat{a}_0 = 0.05 - 0.1$, the studied system sizes turned out to be too small to produce a reliable estimate for $c_1$.

For the lower values of $\hat{a}_0$, the data points of the smaller lattices shown in Fig. 2 begin to show a clear deviation from the straight line behavior. In Fig. 4 we show the raw data for $\hat{R}^2/N_2$ for the four smallest values of $\hat{a}_0$. On the basis of Ansatz (18) this means, that the $c_2$ coefficient, originating from the exponential damping factor of the classical action $S_c$ for the case $\hat{A} \ll 1$, wins more and more. To make this effect more visible and to get more data points for small $\hat{A}$, we performed four additional simulations for the four largest values of $a$ on the lattice sizes as small as $N_0 = 26 - 152$ ($L = 3 - 6$). We now mimic the procedure of Ref. [17] to extract an estimate for $\gamma'_{\text{str}}$ by staying at a rough discretization scale to achieve a sufficiently small $\hat{A}$.

We fitted all data points using Ansatz (18), including a linear and a quadratic term in $1/N_2$, see Table 1. In Ref. [16] it is shown, that in the limit of $\hat{a}_0 \to 0$, $c_1$ should approach $\gamma_{\text{str}} - 2$. Here, because we truncated our fit at some $N_2$, we approach this limit by increasing $a$, and therefore decreasing $\hat{A}$, so that effectively $c_1$ should approach $\gamma'_{\text{str}} - 2$. But, as already mentioned before, this procedure is really ill defined, because it is simply impossible to take the continuum limit of $\hat{a}_0 \to 0$, and staying at the same time at $\hat{A} \ll 1$. We are therefore not convinced, that the coefficient $c_1$ is simply related to $\gamma'_{\text{str}}$.

Nevertheless, by recalling Eqs. (17)-(21) the value of $c_2 \hat{a}_0$ should approach $16\pi^2$. As can be inspected in Table 1, $c_2 \hat{a}_0$ indeed approaches $16\pi^2$, but the total $\chi^2$ of the fit is unacceptably high. The steep increase in the total $\chi^2$ of the four simulations with highest $a$ of course originates in the higher number of data points on the smaller lattice sizes. For those simulations the tail of the power law decay, namely the data points on the large lattices, produce the increase in $\chi^2$. One would expect then that removing the data of the larger lattices would lead to an improvement, because they belong to the regime where $\hat{A}$ is large and the $c_2$ coefficient is not so important. We therefore successively discarded the larger system sizes until we obtained a fit with an acceptable quality. The final values of the fit parameters along with the remaining number of degrees of freedom ($dof$) can be found in Table 2. Indeed we see that we can stabilize the fits much better, and that the value for the classical action $S_c$ comes out very well. Only the simulations with largest $R^2$ coupling $a$ show the first signs of numerical problems we experience with large values of $a$.

Now we have the paradoxical situation that one needs very small lattice sizes or very large values of $a$ to expect good results. On very small system sizes, however, one needs to worry about finite-size effects, which can lead to rather large systematic errors. As a final estimate of this analysis we take the average of the four MC estimates, but due to the aforementioned criticism of this method, we enlarge the error, leading to $\gamma'_{\text{str}} = c_1^{(0)} + 2 = 0.1(3)$ as a conservative estimate. From (6) we see that this value is still consistent with the prediction $\gamma'_{\text{str}} = 0$.

In Ref. [17] a similar value was found for the sphere, but the value of $\gamma'_{\text{str}}$ for the bi-torus ($g = 2$) did not agree with the theoretical expectations. We suggest two possible explanations for this. The first is that the presence of large FSS corrections is responsible for the "failure" of the bi-torus analysis, because the results were obtained on relatively small lattice sizes. The other is that actually the methods presented in Ref. [17] cannot predict anything about $\gamma'_{\text{str}}$, because inherently the continuum limit in $\hat{a}_0$ cannot be taken. Therefore the coincidence of our results with the theoretical value of $\gamma'_{\text{str}}$ for the sphere might be purely accidental.

Still we investigated our data a bit more closely. Another way of improving the quality of the fits is to try out other scaling Ansätze or to include more correction terms to (18). We therefore used (18) augmented by a term $c_3/N_2^3$:

$$\hat{R}^2(\hat{a}_0, N_2)/N_2 = c_0(\hat{a}_0) + c_1(\hat{a}_0)/N_2 + c_2(\hat{a}_0)/N_2^2 + c_3(\hat{a}_0)/N_2^3. \qquad (27)$$

We also tried an Ansatz of $\hat{R}^2(\hat{a}_0, N_2)$ in terms of $\sqrt{N_2}$:

$$\hat{R}^2(\hat{a}_0, N_2)/N_2 = c_0(\hat{a}_0) + c_1(\hat{a}_0)/N_2 + c_{3/2}(\hat{a}_0)/N_2^{3/2} + c_2(\hat{a}_0)/N_2^2 + \ldots \qquad (28)$$

The results for the two generalized fits can be inspected in Tables 3 and 4. We find that both Ansätze seem to work equally good or bad, so that on the basis of these results alone one cannot draw any conclusion on their validity. Actually, the inclusion of more correction terms does not improve the simulations with large and small values of $a$, only the crossover region seems to get improved. Finally one can fit the values of the coefficient $c_1$ obtained according to the Ansätze (18), (28) and (27) versus $1/\hat{a}_0$, see Fig. 5, to make the extrapolation $\hat{a}_0 \longrightarrow 0$ and to see how the observations of Ref. [16] comply with our findings. Fig. 5 was composed using the values from Tables 1, 3, and 4. For small values of $\hat{a}_0$ all three curves seem to come closer together, and are approximately around the theoretical value of $\gamma'_{\text{str}}$. For large values of $\hat{a}_0$ one sees no way how to extract $\gamma_{\text{str}}$. If one extrapolates to large values of $\hat{a}_0$, then this means $\hat{A} \gg 1$, but in the same limit the discretization scale becomes very large, and we can not be sure if we still reach the continuum limit.

To conclude this subsection it should be stressed that at the couplings where we can compare with with Ref. [16], the raw data for $R^2$ do agree within error bars, so that differences in the final results cannot be blamed on using different simulation techniques. It is rather our much larger lattice sizes and the considerably increased range of $\hat{a}_0$ which reveals the potential problems with the approach of Refs. [16, 17].

## 4.2 Scaling at fixed $\hat{A}$

The raw data of our simulations at fixed $\hat{A}$ is shown in Fig. 6. One first notes that all curves of $\hat{R}^2/N_2$ versus $1/N_2$ for different fixed values of $\hat{A}$ are significantly curved. Straight line behavior is visible only asymptotically for large values of $N_2$.

Because of the non-linear FSS behavior observed in Fig. 6, the task of extracting the coefficient $d_1$ of Ansatz (22) proves to be a difficult one. In our attempt to fit all available data points we therefore used besides the FSS Ansatz (22) also a three-parameter fit of the form

$$\hat{R}^2(\hat{A}, N_2)/N_2 = d_0(\hat{A}) + d_1(\hat{A})/N_2 + d_{3/2}(\hat{A})/N_2^{3/2} + \ldots, \qquad (29)$$

which yielded a better fit, with a $\chi^2$ that was mostly only half as high as for the Ansatz (22). Still neither of those fits yielded a satisfactory total $\chi^2$. This can mean that either both Ansätze plainly do not work, or that FSS corrections are still so large that one would need even more correction terms, which, in light of the few available data points, is not applicable. We therefore tried to discard the data on the smaller lattices until we obtained a fit with a reasonable $\chi^2$. The values of the fit parameters together with the total $\chi^2$ and the remaining number of degrees of freedom ($dof$) can be found in Tables 5 and 6. As a general trend one observes that the acceptable fit range increases with decreasing values of $\hat{A}$, with the exception of the two simulation with smallest $\hat{A}$. The first part of the observation can be inferred from Fig. 6, because the curves show more linear behavior the smaller $\hat{A}$ gets. The simulations for the two smallest $\hat{A}$ do not follow this trend because the large values of the $R^2$ coupling $a$ produce numerical problems on the larger lattices. Some configurations seem to freeze in local minima and our algorithm does not seem to be able to relax those minima sufficiently fast.

The resulting plot of both values for $d_1$ versus $\hat{A}$ is shown in Fig. 7. We observe that qualitatively both curves fulfill the theoretical expectations, namely they show a divergence at small $\hat{A}$ and a flattening slope at large $\hat{A}$.

To extract $\gamma_{\mathrm{str}}$ we need to look at large values of $\hat{A}$. Because the divergent term goes like $16\pi^2/\hat{A}$ this means large compared with $16\pi^2$. Looking at the five highest values of $\hat{A}$ we note that no clear linear slope can be observed, so that with the present data it is impossible to obtain a reliable estimate for $\gamma_{\mathrm{str}}$. From a crude linear fit it appears that $\gamma_{\mathrm{str}}$ is too negative, which goes just in the opposite direction of what was claimed in Ref. [16]; however this observation has to be taken with great care. As far as the extraction of $\gamma_{\mathrm{str}}$ is concerned we must thus conclude that either $\hat{A}$ is not yet large enough to see the linear behavior or the presence of large FSS corrections in the first fit have led us to underestimate the errors in our data. The first obstacle could be overcome by increasing $\hat{A}$, whereas the second would require even larger system sizes. Unfortunately it turns out that we need to follow both suggestions simultaneously to improve our data. We have seen in Fig. 6 that with increasing $\hat{A}$ the curves rapidly deviate away from linear behavior already for quite large system sizes. Therefore increasing $\hat{A}$ requires also increasing the system sizes to obtain a comparable accuracy, which in turn requires more and more computing efforts which are beyond the scope of the present study.

The situation for $\gamma'_{\mathrm{str}}$ appears somewhat better. Performing a three-parameter fit of the form (24) to the four data points with $\hat{A} < 16\pi^2$, we obtain $S_c = 399(46)$ and $\gamma'_{\mathrm{str}} = -2.2(1.9)$, if $d_1$ is taken from the fits to the Ansatz (22), and $S_c = 608(77)$ and $\gamma'_{\mathrm{str}} = -6.6(3.1)$, with $d_1$ from Ansatz (29). Already the deviations of the two fit results from each other show that one should not trust these numbers. At least for $S_c$ we know that a value of $16\pi^2 \approx 157.91$ should emerge. This shows that in principle we should include more FSS corrections to stabilize the fits, apart from the obvious need for more data points.

The next thing we tried is to discard in the first fits at fixed $\hat{A}$ the data points for small systems to reach a range where a linear fit of the form

$$\hat{R}^2(\hat{A}, N_2)/N_2 = d_0(\hat{A}) + d_1(\hat{A})/N_2 \qquad (30)$$

is sufficiently accurate. The resulting fits for the six smallest values of $\hat{A}$ are shown in Fig. 8 and the fit data can be found in Table 7. For sufficiently large $N_2$ we see a linear behavior, only the two simulation sets with smallest $\hat{A}$ are somewhat scattered around the fit line. As already mentioned we attribute this to numerical problems in thermalizing the configurations with extremely large $R^2$ coupling, which is presumably not reflected by the computed jack-knife errors. The fit over the resulting six data for $d_1$ according to Eq. (24) is shown in Fig. 9a and yields $S_c = 199(65)$ and $\gamma'_{\rm str} = 5.1(3.0)$, with a reasonable total chi-square of $\chi^2 \approx 3.5$. By looking at the graph, however, one gets the impression as if the four data points with lowest $\hat{A}$ lay more on a scaling curve like Eq. (24), suggesting that the values of $\hat{A} > 158$ are already too large to show the asymptotic behavior. Indeed, these values may already belong to the crossover regime to the weak $R^2$ scaling. In Fig. 9b we therefore fitted just the first four data points, resulting in $S_c = 187(104)$ and $\gamma'_{\rm str} = 5.1(6.6)$, with $\chi^2 = 0.63$. To see what difference it would make to constrain $S_c$ to its theoretical value of $16\pi^2$ we also included a two-parameter fit of the form $d_1 = 16\pi^2/\hat{A} + \gamma'_{\rm str} - 2 + b_0\hat{A}$, that resulted in $\gamma'_{\rm str} = 7(7)$. To the eye both curves are almost identical. Within the relatively large error bars the values of both fits are compatible with the theoretical expectations. In comparison with the analysis in the preceding subsection the estimates appear less accurate, but this time uncontrolled systematic finite-size effects are definitely reduced. To enhance the accuracy we would again need much more data points, now at small values of $\hat{A}$. Here it not necessary to go to very large system sizes in the first fit, but one faces the numerical problem of simulating at extremely large $R^2$ coupling, where the generation of a reliable sample of MC configurations for probing the partition function in (13) proved to be very difficult.

# 5  Conclusions

We tried to measure the string susceptibilities $\gamma_{\rm str}$ and $\gamma'_{\rm str}$ using two FSS Ansätze in $\hat{R}^2$. Although the approach of Ref. [16] is in principle applicable to determine $\gamma_{\rm str}$ correctly, our results for the $dl/l$ measure show that it fails in practice because the system sizes needed in order to reach the thermodynamic (large area) limit $N_2 \longrightarrow \infty$ on reasonable small discretization scales $\hat{a}_0$, are too large. Effectively this can be seen in a crossover behavior from weak to strong $R^2$ scaling, which is mainly due to the fact, that already in the first truncated Ansatz of Eq. (18) one mixes data with small and large $\hat{A}$ by varying $N_2$. Only if $N_2$ could be made sufficiently large, one would be able to stay always in the large $\hat{A}$ regime.

The method of Ref. [17] to estimate $\gamma'_{\rm str}$ with the same Ansatz can be performed practically, but here we encounter severe conceptual problems. The analysis was done employing only very small lattice sizes, and to reach a sufficiently low $\hat{A}$ one has to discard successively the larger lattice systems and to extrapolate to very small system sizes. In this way one will necessarily experience large finite-size effects. Our estimates for $\gamma'_{\rm str}$, obtained for smaller discretization scales and larger lattices, agree with the finding of [17], showing, that finite-size effects seem to be unimportant for the sphere. Nevertheless the continuum limit $\hat{a}_0 \to 0$ can only be performed for some finite value of $N_2$. The coincidence of our value for $c_1^{(0)} + 2$ with $\gamma'_{\rm str}$ might therefore be purely accidental and should not be considered as a significant test of Regge calculus. In light of all these conceptual problems the "failure" of the bi-torus analysis in Ref. [17] should not be regarded as a serious problem for Regge calculus, but rather as a problem with the method itself.

Alternatively, if one works at a well controlled $\hat{A}$ as in (22), one should in principle be able to explore both scaling limits and to predict both $\gamma_{\rm str}$ and $\gamma'_{\rm str}$. We have experienced, however, large FSS corrections in the weak $R^2$ regime, and numerical problems in the strong $R^2$ regime, which make it very difficult to extract the coefficient $d_1$ with high accuracy. Unfortunately, these are the regions in which a precise knowledge of $d_1(\hat{A})$ is needed to extract the asymptotic behavior of $Z(A)$ and the associated string susceptibility exponents. For this reason we were not yet capable to measure either $\gamma_{\rm str}$ or $\gamma'_{\rm str}$ with high enough precision to decide about a "failure" or "non-failure" of Regge quantum gravity. The results we obtained so far are still consistent with the theoretical prediction of Ref. [18] for both $\gamma_{\rm str}$ and $\gamma'_{\rm str}$, although we must admit, that the present data does not provide compelling evidence for either direction. Conceptually, however, already these exploratory results are encouraging and we believe that with higher computational effort one can arrive at a conclusive statement. A particularly challenging test would be the topology of the bi-torus or surfaces of higher gender where the methods of Ref. [16, 17] seem to fail also for $\gamma'_{\text{str}}$.

Still, in the long term, it would of course be desirable to develop more direct approaches to measure $\gamma_{\text{str}}$, as has been done for the DTRS method [1]. There has been one alternative method to test the scaling behavior of Regge gravity [24], where the universal loop length distribution is used. On the basis of a numerical study it is shown, that the $dl/l$ measure gives the theoretically predicted distribution at least for the baby-loop distribution, but only for very large system sizes. Unfortunately it is not yet known, if the scaling of the loop length distribution is directly related to the string susceptibility $\gamma_{\text{str}}$. However, for the uniform $dl$ measure, no agreement was found. This naturally raises again the question of the existence of a correct measure for quantum Regge calculus [15, 23, 25]. Other, more physically motivated measure choices are presently under investigation. We conclude with the main result of our study, that even for the $dl/l$ measure, a failure of Regge calculus to produce the theoretical predictions about the string susceptibilities has not yet been shown.

# Acknowledgments

W.J. gratefully acknowledges a Heisenberg fellowship by the DFG. The numerical simulations were performed on the North German Vector Cluster (NVV) under grant bvpf01 and at the HLRZ in Jülich under grant hbu001.

# References

[1] J. Ambjørn, in proceedings of the 1994 Les Houches Summer School, Session LXII, and references therein;
J. Ambjørn, S. Jain, and G. Thorleifsson, Phys. Lett. B 307 (1993) 34.

[2] V.G Knizhnik, A.M. Polyakov, and A.B. Zamalodchikov, Mod. Phys. Lett. A3 (1988) 819;

F. David, Mod. Phys. Lett. A3 (1988) 1651;
J. Distler and H. Kawai, Nucl. Phys. B321 (1989) 509.

[3] For a recent reviews, see P. Di Francesco, P. Ginsparg, and J. Zinn-Justin, Phys. Rep. 254 (1995) 1;
P. Ginsparg, lectures given at Trieste summer school 1991, LA-UR-91-4101 and hep-th/9112013; published in the proceedings *1991 Summer School in High Energy Physics and Cosmology*, ICTP Series in Theoretical Physics, Vol. 8, edited by E. Gava, K. Narain, S. Randjbar-Daemi, E. Sezgin, and Q. Shafi (World Scientific, Singapore, 1992); p. 785.

[4] V.A. Kazakov, JETP Lett. 44 (1986) 133; Phys. Lett. A119 (1986) 140;
D.V. Boulatov and V.A. Kazakov, Phys. Lett. B186 (1987) 379.

[5] V.A. Kazakov, Phys. Lett. B150 (1985) 182;
F. David, Nucl. Phys. B257 (1985) 543.

[6] J. Jurkiewicz, A. Krzywicki, B. Petersson, and B. Soderberg, Phys. Lett. B213 (1988) 511;
S. Catterall, J. Kogut, and R. Renken, Phys. Rev. D45 (1992) 2957;
C. Baillie and D. Johnston, Mod. Phys. Lett. A7 (1992) 1519;
R. Ben-Av, J. Kinar, and S. Solomon, Int. J. Mod. Phys. C3 (1992) 279.

[7] T. Regge, Nuovo Cimento 19 (1961) 558.

[8] H. Hamber, in proceedings of the 1984 Les Houches Summer School, Session XLIII, edited by K. Osterwalder and R. Stora (North Holland, Amsterdam, 1986); p. 375.

[9] P. Menotti, Nucl. Phys. B (Proc. Suppl.) 17 (1990) 29.

[10] M. Vekić, S. Liu, and H. Hamber, Phys. Lett. B329 (1994) 444; Phys. Rev. D51 (1995) 4287.

[11] W. Beirl and B.A. Berg, Nucl. Phys. B452 (1995) 415.

[12] M. Gross and H. Hamber, Nucl. Phys. B364 (1991) 703.

[13] C. Holm and W. Janke, Phys. Lett. B335 (1994) 143.

[14] C. Holm and W. Janke, Nucl. Phys. B (Proc. Suppl.) 42 (1995) 725.

[15] P. Menotti and P.P. Peirano, Phys. Lett. B353 (1995) 444.

[16] W. Bock and J. Vink, Nucl. Phys. B438 (1995) 320.

[17] W. Bock, Nucl. Phys. B (Proc. Suppl.) 42 (1995) 713.

[18] H. Kawai and R. Nakayama, Phys. Lett. B306 (1993) 224.

[19] H.W. Hamber and R.M. Williams, Nucl. Phys. B248 (1984) 392;
H.W. Hamber, Nucl. Phys. B400 (1993) 347;
and for the entropic approach to the unboundedness problem see,
B.A. Berg, Phys. Lett. B176 (1986) 39;
W. Beirl, E. Gerstenmayer, H. Markum, and J. Riedler, Phys. Rev. D49 (1994) 5231; Nucl. Phys. B (Proc. Suppl.) 30 (1993) 764.

[20] B. DeWitt, in *General Relativity - An Einstein Centenary Survey*, edited by S.W. Hawking and W. Israel (University Press, Cambridge, 1979), and references therein.

[21] A.M. Polyakov, Phys. Lett. B103 (1981) 207.

[22] C. Holm and W. Janke, work in progress.

[23] C. Holm and W. Janke, Nucl. Phys. B (Proc. Suppl.) 42 (1995) 722.

[24] J. Nishimura and M. Oshikawa, Phys. Lett. B338 (1994) 187.

[25] J. Nishimura, Prog. Theor. Phys. 94 (1995) 229.

| $a$ | $c_0$ | $c_1$ | $c_2\hat{a}_0$ | $\chi^2$ |
|---:|---:|---:|---:|---:|
| 0.10 | 0.1093(1) | 1.7(4) | -871(766) | 0.8 |
| 1.25 | 0.2106(1) | 3.1(2) | -26(26) | 0.5 |
| 5.00 | 0.2332(1) | 4.8(2) | 27(11) | 9.0 |
| 10.00 | 0.2391(1) | 4.6(3) | 94(7) | 52.2 |
| 20.00 | 0.2430(1) | 5.9(5) | 104(6) | 12.1 |
| 40.00 | 0.2455(1) | 7.9(7) | 119(4) | 20.0 |
| 80.00 | 0.2485(1) | -0.57(9) | 157.95(5) | 239 |
| 160.00 | 0.2495(1) | -1.00(5) | 158.03(1) | 419 |
| 320.00 | 0.2509(1) | -1.18(9) | 158.00(2) | 122 |
| 640.00 | 0.2525(1) | -1.74(9) | 157.99(1) | 152 |

Table 1: Fit results for the Ansatz $\hat{R}^2/N_2 = c_0 + c_1/N_2 + c_2/N_2^2$. We fitted over all available data points. For the four largest values of $a$ we added 4 runs at system sizes of $N_2 = 48 - 300$ ($L = 3 - 6$). As a measure for the goodness-of-fit we included the total $\chi^2$ of the fit. For small $\hat{A}$ we expect $c_2\hat{a}_0 = 16\pi^2 \approx 157.91$.

| $dof$ | $a$ | $c_0$ | $c_1$ | $c_2 \hat{a}_0$ | $\chi^2$ |
|---|---|---|---|---|---|
| 3 | 80.00 | 0.2530(4) | -2.0(2) | 158.42(6) | 11.4 |
| 5 | 160.00 | 0.2520(2) | -1.7(6) | 158.14(2) | 16.4 |
| 6 | 320.00 | 0.2522(2) | -1.6(1) | 158.04(2) | 17.0 |
| 6 | 640.00 | 0.2535(2) | -2.1(1) | 158.00(1) | 75.3 |

Table 2: Fit results for the Ansatz $\hat{R}^2/N_2 = c_0 + c_1/N_2 + c_2/N_2^2$. We discarded successively the largest lattices until we obtained a reasonable total $\chi^2$ of the fit. The number of degrees of freedom for the fit is denoted by $dof$. For small $\hat{A}$ we expect $c_2 \hat{a}_0 = 16\pi^2 \approx 157.91$.

| $a$ | $c_0$ | $c_1$ | $c_2 \hat{a}_0$ | $c_3$ | $\chi^2$ |
|---|---|---|---|---|---|
| 0.10 | 0.10929(8) | 1.7(8) | -793(5519) | -5146(356559) | 0.8 |
| 1.25 | 0.21063(2) | 3.0(3) | -6(170) | -16999(143587) | 0.5 |
| 5.00 | 0.23309(4) | 5.6(5) | -134(68) | 541437(225053) | 3.2 |
| 10.00 | 0.23877(5) | 8.6(7) | -207(43) | 1913421(269835) | 1.9 |
| 20.00 | 0.24275(9) | 9(1) | -28(40) | 1772617(519739) | 0.4 |
| 40.00 | 0.24519(11) | 13(2) | 6(26) | 2951788(674917) | 0.9 |
| 80.00 | 0.24826(8) | 0.6(2) | 156.4(3) | 10324(1323) | 178 |
| 160.00 | 0.24944(5) | -0.4(1) | 157.7(7) | 4693(758) | 380 |
| 320.00 | 0.25091(7) | -1.1(2) | 157.96(5) | 978(1254) | 122 |
| 640.00 | 0.25253(7) | -2.3(2) | 158.09(3) | -5396(1345) | 136 |

Table 3: Fit results for the Ansatz $\hat{R}^2/N_2 = c_0 + c_1/N_2 + c_2/N_2^2 + c_3/N_2^3$. We fitted over all available data points. As a measure for the goodness-of-fit we included the total $\chi^2$ of the fit. For small $\hat{A}$ we expect $c_2 \hat{a}_0 = 16\pi^2 \approx 157.91$.

| $a$ | $c_0$ | $c_1$ | $c_{3/2}$ | $c_2 \hat{a}_0$ | $\chi^2$ |
|---:|---:|---:|---:|---:|---:|
| 0.10 | 0.1093(2) | 1.5(2.0) | 6(105) | -1288(6981) | 0.81 |
| 1.25 | 0.2106(1) | 2.9(7) | 10(41) | -80(225) | 0.44 |
| 5.00 | 0.2330(1) | 7.6(1.2) | -163(63) | 248(86) | 2.19 |
| 10.00 | 0.2386(1) | 14(2) | -515(77) | 428(50) | 6.65 |
| 20.00 | 0.2426(2) | 15(3) | -519(150) | 279(51) | 0.02 |
| 40.00 | 0.2449(2) | 22(4) | -816(185) | 255(31) | 0.44 |
| 80.00 | 0.2480(1) | 4.1(5) | -98(9) | 160.9(3) | 118 |
| 160.00 | 0.2493(1) | 1.4(3) | -50(6) | 158.8(1) | 337 |
| 320.00 | 0.2509(1) | -0.2(5) | -20(9) | 158.16(7) | 117 |
| 640.00 | 0.2525(1) | -2.5(5) | 16(10) | 157.92(4) | 149 |

Table 4: Fit results for the Ansatz $\hat{R}^2/N_2 = c_0 + c_1/N_2 + c_{3/2}/N_2^{3/2} + c_2/N_2^2$. We fitted over all available data points. As a measure for the goodness-of-fit we included the total $\chi^2$ of the fit. For small $\hat{A}$ we expect $c_2 \hat{a}_0 = 16\pi^2 \approx 157.91$.

| $dof$ | $\hat{A}$ | $d_0$ | $d_1$ | $d_2$ | $\chi^2$ |
|---:|---:|---:|---:|---:|---:|
| 6 | 14.26 | 0.2514(2) | 21(2) | -29143(4579) | 20 |
| 6 | 28.52 | 0.2508(2) | 4(2) | -13173(3462) | 28 |
| 8 | 57.04 | 0.24971(5) | -8.1(4) | 2017(200) | 6.4 |
| 7 | 114.08 | 0.24904(6) | -22.7(7) | 10171(604) | 15 |
| 5 | 228.15 | 0.24830(7) | -47(2) | 51059(6436) | 7.6 |
| 5 | 456.30 | 0.2473(1) | -80(1) | 93358(2356) | 21 |
| 4 | 912.60 | 0.2465(1) | -146(3) | 255536(16586) | 14 |
| 4 | 1825.20 | 0.2450(1) | -234(3) | 428585(16607) | 17 |
| 3 | 3650.40 | 0.2434(1) | -399(5) | 950519(40977) | 15 |
| 3 | 7300.80 | 0.2402(1) | -630(4) | 1698794(31367) | 26 |

Table 5: Fit results for the Ansatz $\hat{R}^2/N_2 = d_0 + d_1/N_2 + d_2/N_2^2$ with $\hat{A}$ kept constant. We discarded successively the smaller system sizes until the fit reached a reasonable total $\chi^2$. The number of the degrees of freedom of the fit is denoted by $dof$.

| $dof$ | $\hat{A}$ | $d_0$ | $d_1$ | $d_{3/2}$ | $\chi^2$ |
|---|---|---|---|---|---|
| 6 | 14.26 | 0.2513(2) | 30(4) | -991(162) | 22 |
| 6 | 28.52 | 0.2507(2) | 9(3) | 516(126) | 26 |
| 8 | 57.04 | 0.24981(6) | -10.7(7) | 150(15) | 3.5 |
| 7 | 114.08 | 0.24917(6) | -28(1) | 502(30) | 8.3 |
| 5 | 228.15 | 0.24845(9) | -58(3) | 1452(181) | 6.2 |
| 5 | 456.30 | 0.2476(1) | -100(3) | 2649(157) | 15 |
| 4 | 912.60 | 0.2469(1) | -184(5) | 5957(384) | 9.8 |
| 4 | 1825.20 | 0.2457(1) | -298(5) | 9992(384) | 6.1 |
| 4 | 3650.40 | 0.2439(2) | -479(5) | 16865(394) | 21 |
| 3 | 7300.80 | 0.2419(2) | -825(7) | 34719(640) | 12 |

Table 6: Fit results for the Ansatz $\hat{R}^2/N_2 = d_0 + d_1/N_2 + d_{3/2}/N_2^{3/2}$ with $\hat{A}$ kept constant. We discarded successively the smaller system sizes until the fit reached a reasonable total $\chi^2$. The number of the degrees of freedom of the fit is denoted by $dof$.

| $dof$ | $\hat{A}$ | $d_0$ | $d_1$ | $\chi^2$ |
|---|---|---|---|---|
| 6 | 14.26 | 0.2522(2) | 13(3) | 61 |
| 6 | 28.52 | 0.2508(2) | 6(2) | 38 |
| 5 | 57.04 | 0.2494(2) | -5(3) | 4.7 |
| 4 | 114.07 | 0.2487(2) | -13(5) | 4.2 |
| 4 | 228.15 | 0.2483(2) | -44(3) | 2.7 |
| 4 | 456.30 | 0.2476(2) | -82(3) | 2.9 |

Table 7: Fit results for the linear Ansatz $\hat{R}^2/N_2 = d_0 + d_1/N_2$ with $\hat{A}$ kept constant for small values of $\hat{A}$. We discarded successively the smaller system sizes until the fit reached a reasonable total $\chi^2$, compare also Fig. 8. The number of the degrees of freedom of the fit is denoted by $dof$.

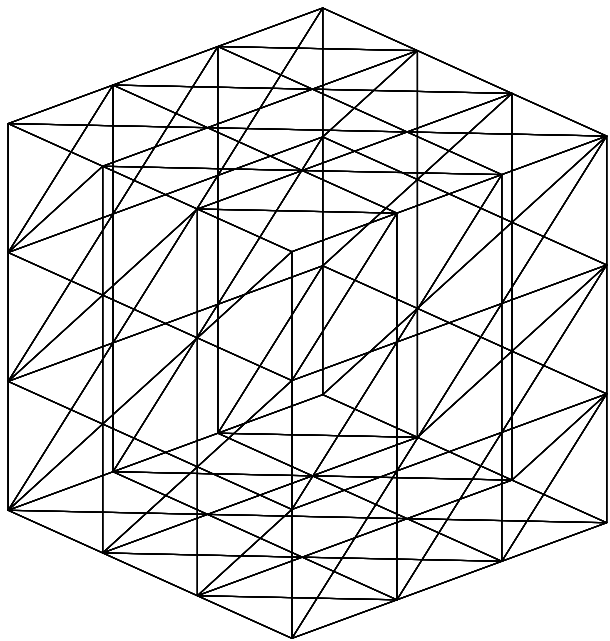

Figure 1: The lattice realization of a spherical topology as the surface of a three dimensional cube with $L = 4$, $N_0 = 56$, and $N_2 = 108$.

Figure 2: Scaling of $\hat{R}^2/N_2$ versus $1/N_2$ for the four lowest values of $1/\hat{a}_0$. Included are the results of the linear fit Ansatz $\hat{R}^2(\hat{a}_0, N_2) = N_2 c_0(\hat{a}_0) + c_1(\hat{a}_0)$.

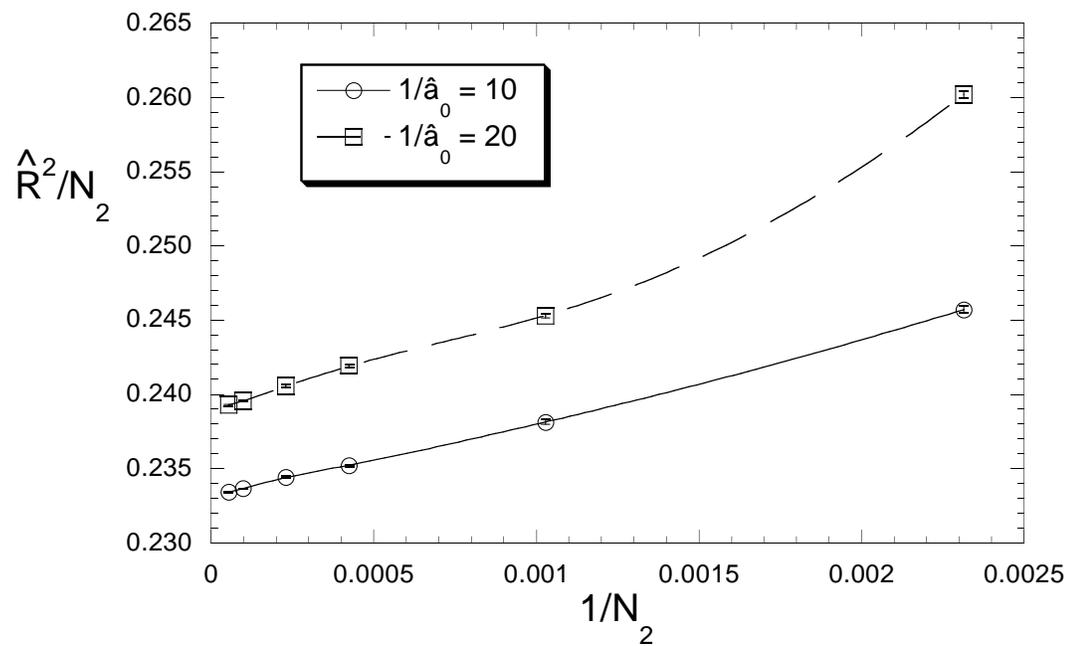

Figure 3: Scaling of $\hat{R}^2/N_2$ versus $1/N_2$ for $1/\hat{a}_0 = 10$ and $1/\hat{a}_0 = 20$. The solid and dashed lines serve as a guide to the eye.

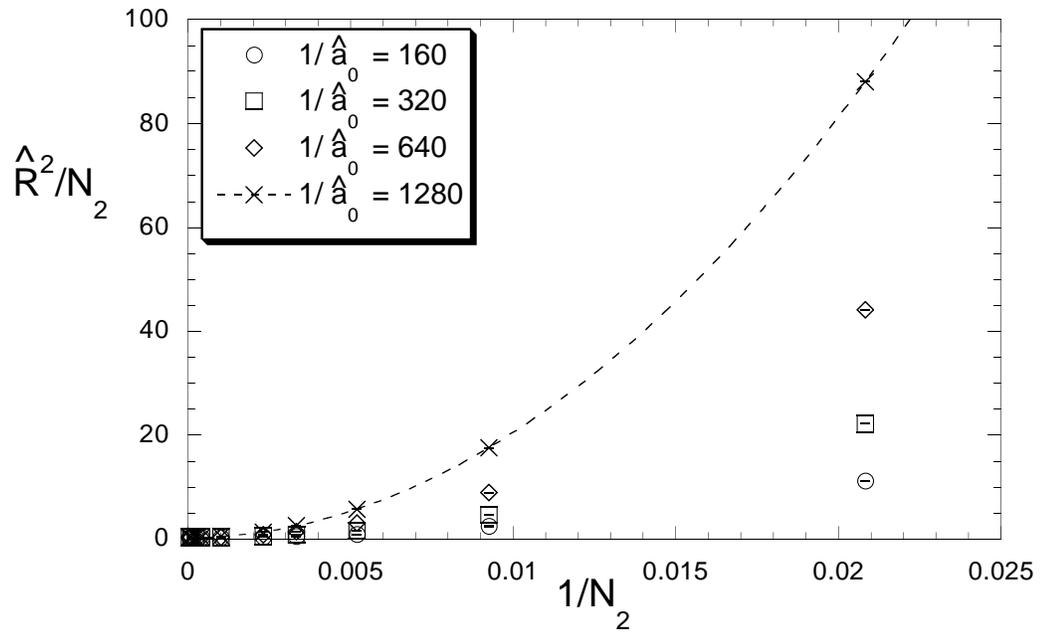

Figure 4: Scaling of $\hat{R}^2/N_2$ versus $1/N_2$ for the four largest values of $1/\hat{a}_0$. Note that the smallest system size was lowered to $N_2 = 48$ ($L = 3$). The dashed curve is the result of a fit of the form (18) over all available data points.

Figure 5: $c_1$ plotted vs. $1/\hat{a}_0$ where $c_1$ was obtained with Eqs. (18), (28), and (27). The two solid lines show the theoretical expectations for $\gamma_{\text{str}}$ and $\gamma'_{\text{str}}$.

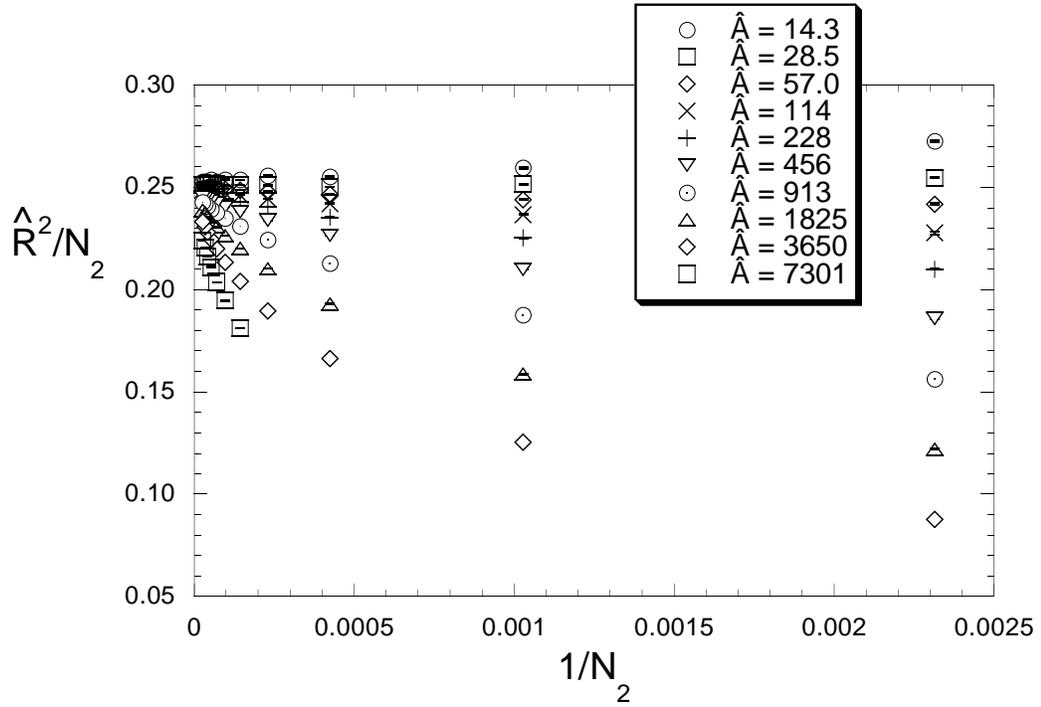

Figure 6: Scaling of $\hat{R}^2/N_2$ versus $1/N_2$ for all simulations with constant $\hat{A} = 9126/a$, where $\hat{A}$ ranges from 14 to 7300.

Figure 7: $d_1$ versus $\hat{A}$, where $d_1$ was extracted according to the two different finite-size scaling Ansätze (22) and (29), showing all available data points.

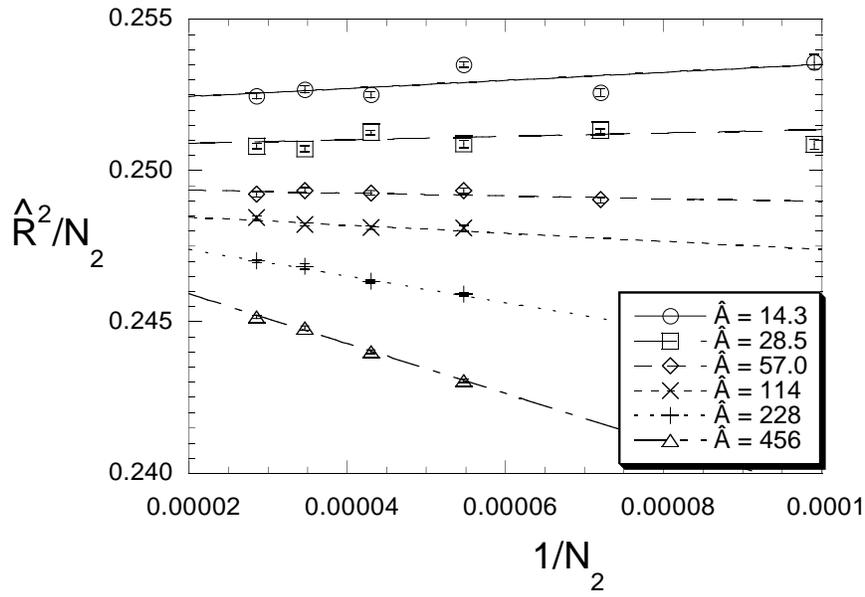

Figure 8: Scaling of $\hat{R}^2/N_2$ versus $1/N_2$ for simulations with constant $\hat{A} = 9126/a$, where $\hat{A}$ ranges from 14 to 456. The straight lines denote the linear fits according to Ansatz (30). Only the visible data points were used for the fits.

Figure 9: $d_1$ versus $\hat{A}$, where $d_1$ was extracted according to Ansatz (30), compare Fig. 8. The solid lines show three-parameter fits of the form $S_c/\hat{A} + \gamma'_{\text{str}} - 2 + b_0\hat{A}$ over the six (a) respectively four (b) visible data points. The dashed curve in (b) is a constrained two-parameter fit with $S_c = 16\pi^2$.